# Top Quark Physics – a Popular Review

**Bodo Lampe**

Max-Planck-Institut für Physik

Werner-Heisenberg-Institut

Föhringer Ring 6

D-80805 München

## ABSTRACT

The top quark has been discovered at FERMILAB last year. The following features of top quark physics will be discussed in this article:

- the top quark in the standard model

- production and decay of the top quark in proton collisions
  (direct evidence for top)

- virtual effects of top quarks in electroweak observables
  (indirect evidence for top)

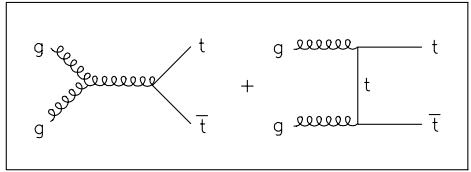

Fig. 1

## 1. Introduction

The top quark has become one of the major interests of high energy particle physicists today [1]. The approval of the "Large Hadron Collider" (LHC) by the CERN member states guarantees that it will remain so over the next 20 years. The LHC [2] is a proton-proton collider experiment with a beam energy of about 8 TeV (1 TeV= $10^{12}$ eV), in which top quarks are produced via the parton subprocesses shown in fig. 1. The incoming gluons (g) shown in fig. 1 are assumed to be composites of the high energetic protons. In fact, for a precise study of top quark properties the start-up of the LHC is indespensible. No other machine will be able to produce nearly as many top quarks. More than 1000000 top quarks are to be expected per year, thus providing an excellent statistics. So the LHC is certainly important for the understanding of the top quark. But how important is the top quark? I will not answer this question directly, but the reader is asked to draw his own conclusions from what follows in this article.

The present knowledge about the top quark mainly comes from two sources. One is the Tevatron [3] at FERMILAB in Chicago, a proton-proton collider with a beam energy of about 900 GeV (1 GeV= $10^9$ eV), in which about 100 top quarks have been produced in the last years. A top quark mass value between 170 and 200 GeV has been reported by the two Tevatron detectors CDF and D0 [3]. The other source is the LEP experiment at CERN in Geneva [4], an $e^+e^-$–annihilation experiment in which indirect evidence for the top quark has been obtained in recent years. Although the beam energy at LEP is not high enough to produce real top quarks, the high statistics collected at the Z–resonance allows to detect higher order effects related to the existence of the top quark. The precision of the LEP experiment is in fact so high that top quark mass values of the same accuracy as from Tevatron have been deducted.

In the standard model of elementary particles the role of the top quark is completely fixed by its position in the spectrum of the spin–$\frac{1}{2}$ particles. This can be seen in fig. 2, where the leptons are placed in the left and the quarks in the right column of the table. The three rows correspond to the three families of elementary fermions kwown today. The quarks of the first family constitute the ordinary matter (p,n,...), whereas the second family is needed for strange and charmed particles (K,D,...). All lefthanded fermions appear as weak isospin doublets, whereas all right handed quarks are singlets. This is related to the parity violating nature of the weak interaction, which do not act on the righthanded components. The top quark appears as the weak isospin partner of the bottom quark and is necessary to complete the spectrum of the third family. This was realized soon after the discovery of the bottom flavour in 1977 and it was subsequently believed that the mass of the top quark should be somewhere between 10 and – at most – 50 GeV. As time went by, the accelarators reached these energies; but since the top quark was not found, the lower limit shifted to higher and higher values, until finally last year it was discovered.

In my opinion the large value of the top quark mass $m_t$ is a very interesting property, whose impact on particle physics has not been completely realized. The top quark is as heavy as a heavy nucleus and still – at least within the standard model – a point particle. Since it is so heavy, it is conceivable that deviations from the standard model will first be found in studying top quark properties. The neighbourhood of the top quark to new physics is illustrated in fig. 3, in which the mass and energy scales of the physical universe are drawn on a logarithmic scale. One can associate a characteristic energy scale to each of the fundamental interactions. These are shown to the right of fig. 3. For example, the characteristic scale for the theory of the strong interaction is the so called Λ–scale $M_{QCD}$, of order 1 GeV. It can be interpreted as the scale, above which the strong interaction becomes so weak that perturbation theory can be applied. The characteristic scale for the weak interaction is the Fermi scale $M_F$, about 100 GeV and corresponding roughly to the masses of W and Z. $M_{GUT}$ is the scale of the grand unified theories. At energies $M_{GUT}$ the running (=energy dependent) couplings of the strong and electroweak interactions converge to one value. Within the ordinary standard model the convergence is not very precise. Many of the theoretical colleges believe that an additional new interaction with a characteristic scale $M_{NP}$ of order 1 TeV could clean up this and other ugly features of the standard model. Finally there is the Planck scale which is the characteristic scale for gravity. Particle masses are shown to the left of fig. 3, starting with the massless particles (neutrinos and photons) and ending with the top quark. In intermediate places one finds the first and second family matter and the heavy vector bosons. The important point to notice about fig. 3 is the neighbourhood of $m_t$ to the new physics scale $M_{NP}$. It is true that nobody can say, whether $M_{NP}$ is 1, 10 or 100 TeV, but there is at least a certain chance, that experiments involving the top quark get a first glimpse of it.

| $\begin{pmatrix} \nu_e \\ e^- \end{pmatrix}_L \quad e^-_R$ | $\begin{pmatrix} u \\ d \end{pmatrix}_L \quad \begin{matrix} u_R \\ d_R \end{matrix}$ | —— p,n,π,... |
|---|---|---|
| $\begin{pmatrix} \nu_\mu \\ \mu^- \end{pmatrix}_L \quad \mu^-_R$ | $\begin{pmatrix} c \\ s \end{pmatrix}_L \quad \begin{matrix} c_R \\ s_R \end{matrix}$ | —— K,... |
| $\begin{pmatrix} \nu_\tau \\ \tau^- \end{pmatrix}_L \quad \tau^-_R$ | $\begin{pmatrix} t \\ b \end{pmatrix}_L \quad \begin{matrix} t_R \\ b_R \end{matrix}$ | |

Fig. 2

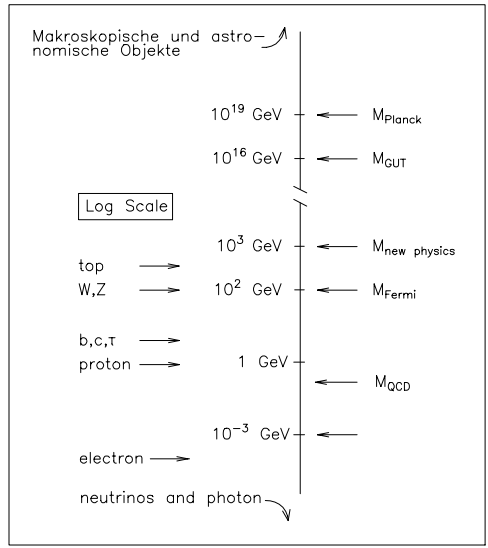

Fig. 3

There are several other important consequences of the large top quark mass, like its extremely short lifetime, to be discussed in the following sections.

## 2. The Top Quark in the Standard Model

I have already stressed that all properties of the top quark except for its mass are fixed by its position in fig. 2. The statement is that the top quark behaves like the corresponding members of the first and second family, the up and the charm quark. More specifically, all couplings of the top quark to photons,gluons,W and

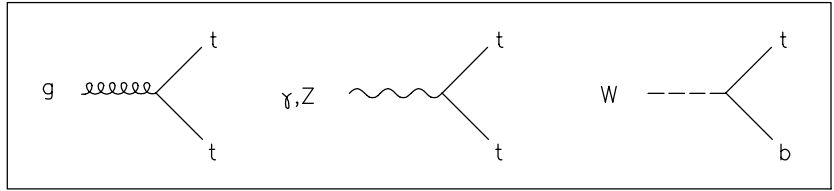

Fig. 4

Z – the mediating particles of the strong and electroweak interactions – are fixed by this statement. In fig. 4 the Feynman diagrams are shown which encode these interactions. The coupling to the gluon is the ordinary QCD vector coupling with strength $g_s$ and induces the production mechanism fig. 1 of top quarks in proton collisions. Correspondingly, the coupling to photon and Z will allow to produce top quarks in $e^+e^-$ annihilation provided an $e^+e^-$ energy of $2m_t$ can be reached. This is the threshold condition needed to get two real top quarks.

The coupling of the top quark to the W is particularly interesting because it induces essentially all decays of the top quark, $t \to W^+ b$. There are mixing effects between the families so that in principle $t \to W^+ s$ is allowed, too, but these are really tiny effects, so that in 99.8% of all cases one will have $t \to W^+ b$. This decay is in principle an ordinary V–A decay like, for example, muon decay. The main difference is that the top quark is so heavy that the W is produced on-shell, i.e. as a real particle. This has a number of consequences for the kinematics of the decay, distributions of decay products etc. It also affects the total width of the top quark (the inverse lifetime) which instead of with the fifth power of $m_t$ increases with the third power. This power behaviour is still so strong that a width of about 1.5 GeV arises for a top quark mass of 175 GeV [5]. It corresponds to an extremely short lifetime of the top quark of about $10^{-24}s$ which is of the order of the lifetime of the W and Z and much

shorter than the lifetimes of all other strongly interacting particles. [1]

The short lifetime of the top quark has an interesting consequence for its strong interaction. It is true that the top quark is a coloured particle just like all the other quarks. Still it does not form baryonic or mesonic bound states, simply because it decays before the bound states are formed. In QCD the formation of bound states takes a certain time $\frac{1}{M_{QCD}} \sim 10^{-23}s$. The top quark decays $10^{-24}s$ after its formation, so that bound states cannot be formed. It should be stressed that the top quark does interact with gluons, but these interactions at time scales less than $10^{-24}s$ are so weak that the top quark exists only as a free quark. This is a very interesting property because it allows to make precise predictions for the top quark in the perturbative sector of the standard model, without all the ambiguities from the perturbative domain. Effects from new physics should show up very clearly on top of it.

As we have argued the property of the top quark not forming bound states is related to the fact that its width is larger than $M_{QCD}$. On the other hand, the width is still much smaller than the mass of the top quark, so that the top quark behaves as a real particle and does not disappear as a flat broad resonance.

## 3. Top Quark Production in Proton Collisions

It was noted earlier that the main production mechanism for the top quark at LHC will be the gluon fusion processes fig. 1. The Tevatron is a proton collider,too ,but at the Tevatron the main production mechanism for the top quark is the light quark annihilation process fig. 6. In the following I want to take some time to explain the difference, because it is the origin for a difference in cross section between Tevatron and LHC of almost three orders of magnitude.

---

[1] The correspondence between mass, energy, time and length is given by $1 GeV^{-1} = 6.6 \, 10^{-25}s = 2 \, 10^{-14} cm = 0.56 \, 10^{27} kg$.

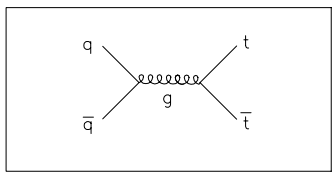

Fig. 6

In today's understanding a high energetic proton consists of massless "partons", i.e. light quarks and gluons which move as collinear quasi-free particles. Interactions with a large momentum transfer between two high energetic protons can be understood as interactions between their partonic constituents, c.f. figs. 1 and 6. Fig. 7 shows a qualitative picture of the number density of light quarks and gluons as a function of their fraction $x = \frac{E_{parton}}{E_{proton}}$ of proton energy. As can be seen, the gluons dominate the small x region, so that a low energy parton in a high enery proton is most probably a gluon. For kinematical reasons two colliding partons need to have an energy of at least $2m_t$ to produce a top quark pair. Due to the high energy protons, at LHC this condition is fulfilled already for quite small values of $x > \frac{350}{16000} \sim 0.02$ whereas at Tevatron $x > 0.2$. Therefore, top production at the Tevatron misses the small x region in which the gluon initiated processes have a large cross section. On the other hand, at LHC one is in the small x region with its overwhelming gluon density.

The quantitative consequences of these qualitative explanations are shown in fig. 8, where the production cross sections for top quarks at Tevatron and LHC are drawn as a function of the top quark mass. One sees that there is a difference of almost three orders of magnitude which makes the LHC extremely superior to the Tevatron. In fact, one should consider this as one of the major propaganda plots for the LHC. Also shown is the expectation for the so called SSC project which was cancelled last year. The SSC was intended to become a proton collider with a beam energy of 2x20 TeV but was stopped by the US government for financial reasons. We see from fig. 8 that this decision is not as desastrous for the top quark as it is for the Higgs particle because the main step in cross section is from the Tevatron to the LHC.

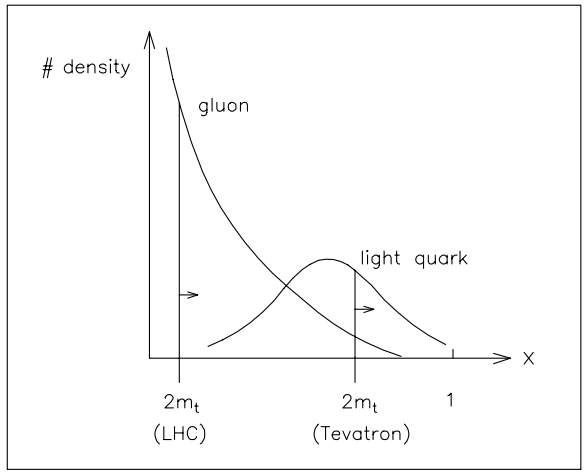

Fig. 7

Once the top quark is produced it has to be detected and its properties have to be determined. Due to its short lifetime it cannot be detected directly but only via its decay products, the b quark and the W boson. b quarks can be detected more or less directly, using vertex detectors, where the trace of the B mesons can be reconstructed from their decay vertex a few mm away from the proton collision point. W's have a lifetime as short as the top quark so that they cannot be detected directly but have to be reconstructed from their decay products, either a lepton plus a neutrino or two light quarks. The leptons from W decays are always high energetic and a hard lepton in conjunction with a high energy b quark jet is, in fact, a rather unambiguous signal for a top quark.

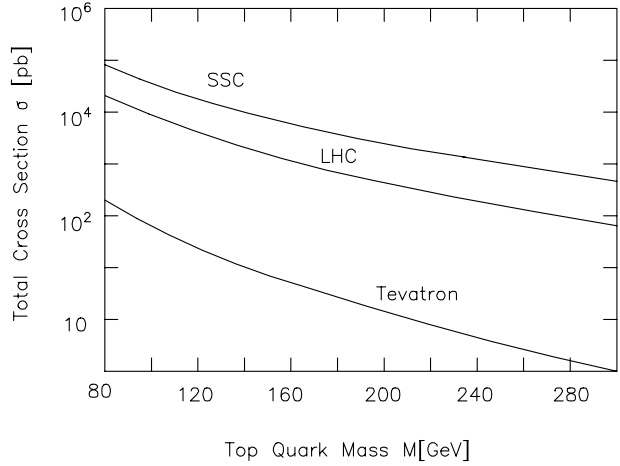

Fig. 8

The top quark mass can be reconstructed from the energy and momenta of its decay products. Consider, for example, an event like is depicted in fig. 9. A $t\bar{t}$ pair is produced and decays subsequently. One of the W's in fig. 9 decays leptonically, the other one hadronically. Events like this are well suited for the determination of the top quark mass, because the appearance of a hard lepton together with b quarks can be used as a trigger to show that a top quark has been produced, and the mass value can be deducted from the on shell mass condition

$$(q_1 + q_2 + p_b)^2 = m_t^2$$

where $p_1$ and $p_2$ are the 4-momenta from the light quark jets from the W decay and $p_b$ is the reconstructed 4-momentum of the b quark. The above equation corresponds to the resonance condition $p_t^2 = m_t^2$ which arises from the Breit–Wigner factor $\frac{1}{(p_t^2 - m_t^2)^2 + m_t^2 \Gamma_t^2}$ which is always present in the cross section.

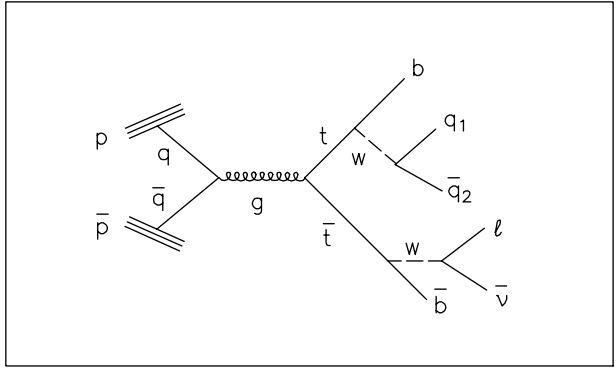

Fig. 9

The main source of error in the mass determination arises, because it is often impossible to distinguish b and $\bar{b}$ so that an ambiguity arises in the determination of $p_b$.

The semileptonic decay of the $\bar{t}$ in fig. 9 is not so well suited for the mass determination because the neutrino escapes the detection so that the momentum of the $\bar{t}$ cannot be fully reconstructed.

In fig. 10b one of the actual events of the type fig. 9 is shown. It can be seen in fig. 10a that the detector corresponds to a surface around the proton beams. This surface is unfolded to a plane in fig. 10b. The energies of the various $t\bar{t}$ decay products are drawn orthogonal to the plane.

## 4. Top Quark Decay

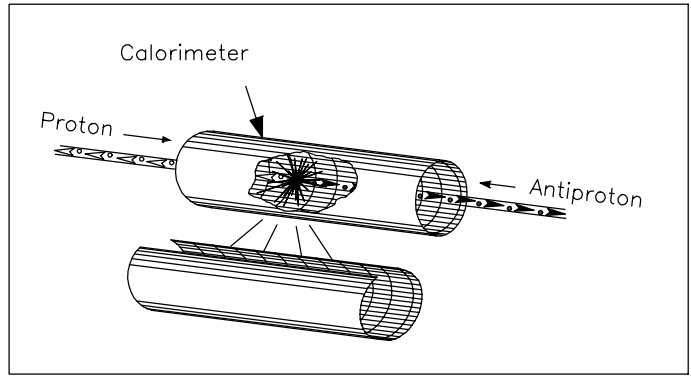

Fig. 10a

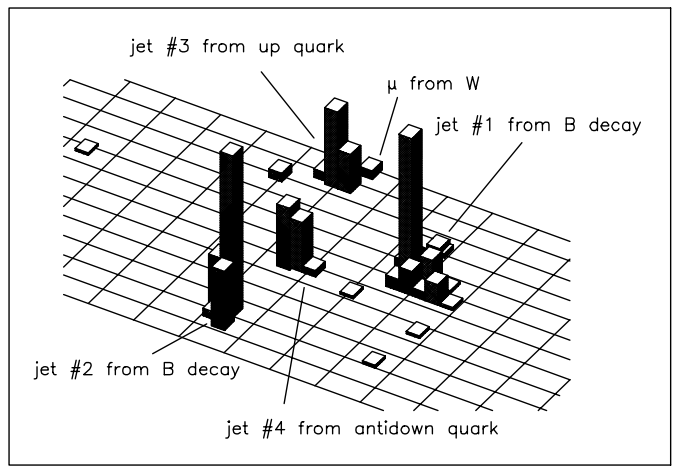

Fig. 10b

Some general features of top quark decay – like the total width and the main signatures – have already been discussed in the previous sections. It has been shown how interesting the large width is from a theoretical point of view and how the knowledge about the decay mechanism can be used to determine the top quark mass. It should be stressed that a lot more can be learned from a careful study of top quark decays ,in particular at the LHC with its excellent statistics. Namely, one can look at certain distributions of decay products which are sensitive to the dynamics of the decay. As was shown in section 2, the decay is in principle an ordinary "lefthanded" V–A decay. It may be worthwhile to make a careful experimental study, though, because some of the particles involved (t and W) are so heavy that deviations from the standard model are conceivable. Interesting distributions are, for example, energy distributions of decay products looked at in the rest system of the top quark. Another interesting example is the distribution in the angle $\theta$ between heavy and light quark direction defined in the rest system of the W, c.f. fig. 11. This distribution has an expansion in powers $\cos\theta$ whose coefficients give information about the contributions from the various W polarisations.

$$\frac{d\Gamma}{d\cos\theta} = \frac{3}{8}\Gamma_T(1+cos^2\theta) + \frac{3}{4}\Gamma_L sin^2\theta$$

It can be shown that there is an appreciable contribution $\Gamma_L$ from longitudinal W's. Relative to the transverse W's they appear in top quark decays at a rate [6]

$$\frac{\Gamma_L}{\Gamma_T} = \frac{m_t^2}{2m_W^2} \sim 2.4$$

It would be very valuable to check this relation by studying the above mentioned angular distribution, in particular in the light of recent results from the LEP experiment where the coupling of the top quark to longitudinal W's arises as a tiny higher order effect. Here, in the decay of the top quark, it is a large first order effect which can be precisely studied.

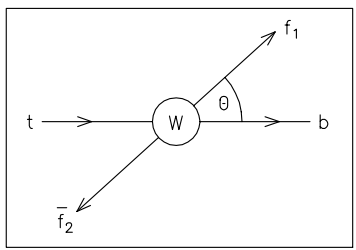

Fig. 11

## 5. The Top Quark as a Higher Order Effect at LEP

The basic process studied at LEP is the production of light fermions on the Z resonance [4], $e^+e^- \to f\bar{f}$, shown in fig.12. In fig. 13 the cross section is given as a function of the $e^+e^-$ beam energy. One can see that at an energy corresponding to the Z mass the cross section is much enhanced due to the Breit–Wigner resonance factor for the Z, and a high statistics experiment can be carried out.

A priori the processes at LEP have nothing to do with the top quark. The top quark enters only as a tiny higher order effect, in the form of the diagrams shown in figs. 14 and 15, but due to the high statistics of the experiment the effects from the top quark can be felt. The top quarks arising in figs. 14 and 15 are not real but only virtual particles far off their mass shell, because the LEP energy is not sufficient to produce real top quarks. The relative magnitude of these diagrams with respect to the leading term fig. 12 is about 1%. Fig. 14 is important for all light fermions whereas fig. 15 arises only in case of b production. Nota bene, that fig. 15 involves the same tbW vertex which also induces the decay of real top quarks.

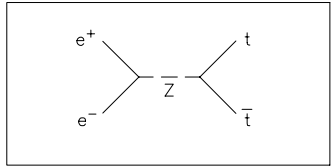

Fig. 12

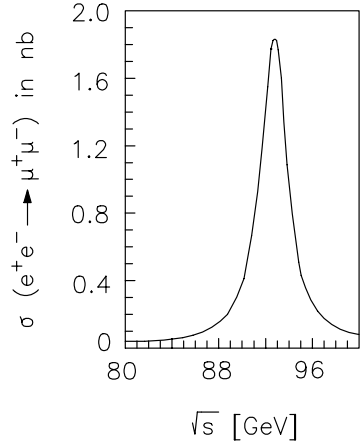

Fig. 13

There are a lot of observables at LEP in which the effect from fig. 14 can be felt. For example, fig. 14 has an effect on the width $\Gamma_Z$ of the Z boson, i.e. the width of the peak shown in fig. 13. In fig. 16 the weak, quadratic $m_t$ dependence of $\Gamma_Z$ is shown as predicted by the theory. According to the LEP measurements the width is given by $\Gamma_Z = 2.50 \pm 0.004$. Using fig. 16 this can be translated into a prediction for the top quark mass, $m_t = 160 \pm 15 \pm 15 GeV$. The first error is due to the experimental error on $\Gamma_Z$. The second error is due to uncertainties in the value of the strong coupling constant $g_s$ and the mass $m_H$ of the Higgs boson. Namely, there are not only higher order effects due to the top quark but also due to virtual gluon exchange and due to the Higgs particle. The Higgs field is an ingredient of the standard model which has not been discovered so far. It is needed for theoretical reasons to introduce the particle masses in a consistent way. The mass of the Higgs field can be anywhere between 60 and 1000 GeV. The dependence on $m_H$ and on $g_s$ is even smaller than on $m_t$, but for a determination of $m_t$ they should be taken into account. If $\Gamma_Z$, $g_s$ and $m_t$ would be known with arbitrary precision, one could use the LEP measurements to determine $m_H$. We shall say more on the possibility of determining the Higgs mass later.

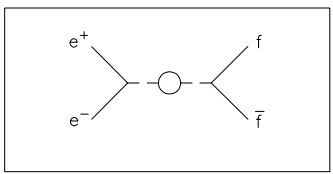

Fig. 14

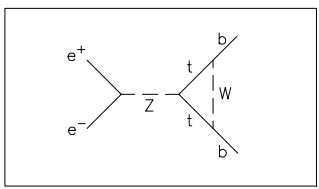

Fig. 15

The LEP way of determining the top quark mass is very indirect and is much more sensitive to effects from new physics than the kinematic $m_t$ determination by the Tevatron. For example, new particles might run in loops similar to the loop fig. 14 and could in principle affect the result much stronger than $g_s$ does. The fact that the two $m_t$ values, $m_t^{LEP}$ and $m_t^{CDF}$ roughly agree, can be taken as a hint that new physics effects should be small. This is true, however, only to the energy scale one is working with, here $m_Z$. New physics effects have the tendency to increase quadratically with energy, so that effects which are tiny at LEP become detectable at LHC!

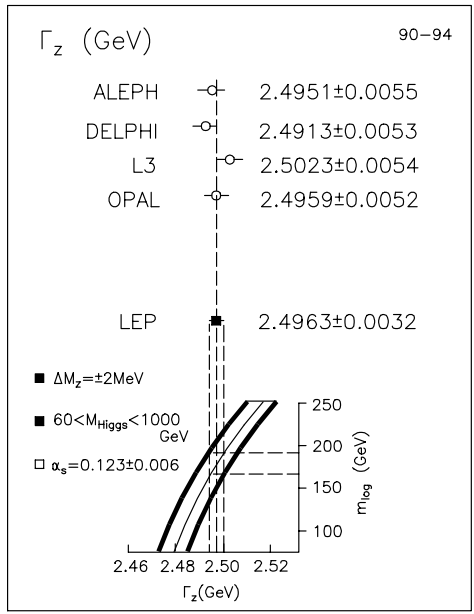

Fig. 16

The quoted value of $m_t^{LEP}$ arises solely from effects of diagram fig. 14. In contrast, the question of loop effects in b quark production (fig. 15) has not yet been settled experimentally. At the moment the experimental analysis of b quark production points to top quark values which are nearer to 120 than to 170 GeV. If this result would persist, this could point to a new physics effect. However, for this analysis an identification of b quarks to per mille accuracy is necessary. Experimentally this seems to be extremely difficult and it is well possible that the explanation of the present discrepancy lies in the misidentification of b and c quarks.

Let us come back to the possibility of determining the Higgs mass through tiny higher order effects. Depending on their mass it is well possible that real Higgs particles will be produced only at later stages of the LHC experiment, if at all. In this case it is an interesting exercise to try to infer $m_H$ from earlier available data. In fact there is a plan to upgrade the Tevatron so as to reach higher beam energies and correspondingly produce more top quarks, of the order of $10^{3-4}$ around the year 2000. That way the experimental error on $m_t$ will become much smaller and it will not be necessary to refer to the LEP analysis of the top quark mass. Instead, the procedure to estimate $m_H$ will be the following:

- Take the most precisely known electroweak parameters. This will be $m_Z$ (from the LEP experiment), the Fermi constant $G_\mu$ (from muon decay) and the fine structure constant $\alpha$. In leading order they allow to determine the W mass via

$$(1 - \frac{m_W^2}{m_Z^2})G_\mu m_W^2 = \frac{\pi\alpha}{\sqrt{2}}$$

- Higher order corrections disturb this relation in a way sensitive to $m_t$ and $m_H$. Thus, for fixed $m_H$ a curve in the $m_W$–$m_t$ plane arises. These curves are shown in fig. 17. They are very sensitive to $m_W$ and much less senitive to $m_t$ and $m_H$ because these are higher order effects. Also shown are the present errors on $m_W$ and $m_t$. These errors are input from present Tevatron data and to be improved after the upgrade.

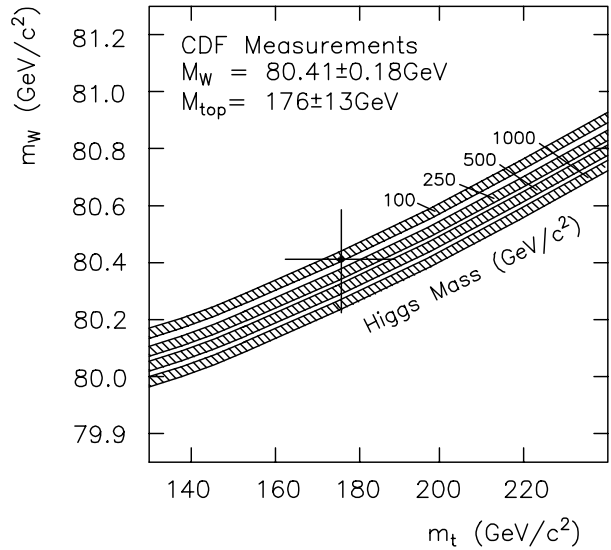

Fig. 17

I think that this figure makes it sufficiently clear that an estimate of the Higgs mass will be possible once the errors on $m_W$ and $m_t$ have decreased.